\documentclass[conference]{IEEEtran}
\IEEEoverridecommandlockouts
\usepackage{cite}
\usepackage{amsmath,amssymb,amsfonts}
\usepackage{algorithmic}
\usepackage{graphicx}
\usepackage{textcomp}
\usepackage{xcolor}
\usepackage{algorithm}
\usepackage{array}
\usepackage[caption=false,font=normalsize,labelfont=scriptsize,textfont=scriptsize]{subfig}
\usepackage{stfloats}
\usepackage{url}
\usepackage{verbatim}
\usepackage{subfig}
\usepackage{amssymb}
\usepackage{siunitx}
\usepackage{threeparttable}
\usepackage{bm}
\usepackage{multirow}
\usepackage{authblk}
\def\BibTeX{{\rm B\kern-.05em{\sc i\kern-.025em b}\kern-.08em
    T\kern-.1667em\lower.7ex\hbox{E}\kern-.125emX}}
\begin{document}

\title{Weighted Codebook Scheme for RIS-Assisted Point-to-Point MIMO Communications}

\author{Zhiheng Yu, Jiancheng An,~\IEEEmembership{Member,~IEEE,} Lu Gan,~\IEEEmembership{Member,~IEEE,} \\Hongbin Li,~\IEEEmembership{Fellow,~IEEE,} and Symeon Chatzinotas,~\IEEEmembership{Fellow,~IEEE}
\thanks{This work was partially supported by National Natural Science Foundation of China 62471096. The work of Hongbin Li was supported in part by the National Science Foundation under Grants CCF-2316865, ECCS-2212940, and ECCS-2332534.}
\thanks{Z. Yu and L. Gan are with the School of Information and Communication Engineering, University of Electronic Science and Technology of China (UESTC), Chengdu 611731, China, and also with the Yibin Institute of UESTC, Yibin 644000, China (email: zhihengyu2000@163.com; ganlu@uestc.edu.cn).}
\thanks{J. An is with the School of Electrical and Electronics Engineering, Nanyang Technological University, Singapore 639798 (e-mail: jiancheng\underline{~}an@163.com).}
\thanks{H. Li is with the Department of Electrical and Computer Engineering, Stevens Institute of Technology, Hoboken, NJ 07030, USA (e-mail: Hong-bin.Li@stevens.edu).}
\thanks{S. Chatzinotas is with the Interdisciplinary Centre for Security, Reliability, and Trust (SnT), University of Luxembourg, 1855 Luxembourg City, Luxembourg and with the Department of Electronic Engineering Kyung Hee University 1732, Deogyeong-daero, Giheung-gu, Yongin-si, Gyeonggi-do, 17104, Korea (e-mail: symeon.chatzinotas@uni.lu).}}

\markboth{DRAFT}%
{Shell \MakeLowercase{\textit{et al.}}: A Sample Article Using IEEEtran.cls for IEEE Journals}

\maketitle

\setlength{\abovedisplayskip}{3pt}  
\setlength{\belowdisplayskip}{3pt}  
\setlength{\floatsep}{0pt}         
\setlength{\textfloatsep}{0pt}     
\setlength{\intextsep}{0pt} 

\begin{abstract}
Reconfigurable intelligent surfaces (RIS) can reshape the characteristics of wireless channels by intelligently regulating the phase shifts of reflecting elements. Recently, various codebook schemes have been utilized to optimize the reflection coefficients (RCs); however, the selection of the optimal codeword is usually obtained by evaluating a metric of interest. In this letter, we propose a novel weighted design on the discrete Fourier transform (DFT) codebook to obtain the optimal RCs for RIS-assisted point-to-point multiple-input multiple-output (MIMO) systems. Specifically, we first introduce a channel training protocol where we configure the RIS RCs using the DFT codebook to obtain a set of observations through the uplink training process. Secondly, based on these observed samples, the Lagrange multiplier method is utilized to optimize the weights in an iterative manner, which could result in a higher channel capacity for assisting in the downlink data transmission. Thirdly, we investigate the effect of different codeword configuration orders on system performance and design an efficient codeword configuration method based on statistical channel state information (CSI). Finally, numerical simulations are provided to demonstrate the performance of the proposed scheme.
\end{abstract}

\section{Introduction}
\IEEEPARstart{R}{econfigurable} intelligent surface (RIS) is a transformative technology in the field of wireless communications \cite{an2021low,yao2025optimizing}. An RIS is a surface embedded with numerous passive elements capable of adjusting their electromagnetic properties dynamically. These elements can intelligently reflect and manipulate radio-frequency (RF) signals during propagation, thereby improving signal strength, reducing interference, or enhancing spectral efficiency \cite{shi2024ris}. Therefore, RIS holds significant potential for various applications by revolutionizing wireless network design \cite{an2024stacked}.

In RIS-aided communications, pivotal challenges encompass channel estimation and passive beamforming. Channel estimation is instrumental for acquiring accurate channel state information (CSI) \cite{wei2021channel}, while passive beamforming is to intelligently control reflection coefficients (RCs) of the RIS to achieve the desired quality of service (QoS) \cite{xu2023antenna}. In previous research, both the instantaneous and statistical CSI have been employed to optimize the RIS RCs. For instance, \emph{Wu et al.}  proposed a dynamic RIS beamforming framework utilizing instantaneous CSI to boost the sum rate of an RIS-aided wireless powered communication network \cite{wu2021irs}. Meanwhile, \emph{Gunasinghe et al.} leveraged statistical CSI to optimize phase shift for multi-cell RIS-aided multi-user massive MIMO systems \cite{gunasinghe2023achievable}.

Nevertheless, the passive beamforming schemes entail excessive pilot overhead. To address this issue, methods with limited feedback have been developed to facilitate the adjustment of phase shifts of the RIS \cite{an2022codebook}. Specifically, \emph{An et al.} proposed a low-complexity framework to maximize the achievable rate, where they analyzed the theoretical performance of random and uniform codebooks \cite{an2021low,an2022Joint}. Moreover, to handle different channel conditions, \emph{Jia et al.} proposed an environment-aware codebook design that leverages the statistical CSI \cite{jia2023environment}, while \emph{Yu et al.} evaluated its theoretical performance in the presence of imperfect CSI \cite{yu2024environment}. As the above codebooks only consider the far-field scenario, \emph{Lv et al.} proposed angular-domain and distance-based codebooks for different channel models in the near-field scenario \cite{lv2023ris}. Additionally, \emph{Wang et al.} adopted the DFT codebook for realizing near-field passive beamforming and proposed a ring-type codebook to achieve signal-to-noise ratio (SNR) enhancement \cite{wang2022ring, TWC_2025_An_Flexible}.

\begin{table}
\scriptsize
\centering
\caption{Comparison of the Proposed Scheme with Previous Approaches} \label{table:1}
\resizebox{8.8cm}{!}{
\begin{tabular}{|c|c|c|c|c|} 
\hline
Scheme & Reference & Overhead & Complexity & RIS RC solution \\ 
\hline
CE \& PBF & \cite{wu2021irs,kundu2022optimal,gunasinghe2023achievable} & $N+1$ & High & $\boldsymbol{\varphi} \in \mathbb{C}^N$\\
\hline
PC & \cite{an2021low,an2022codebook,an2022Joint,jia2023environment,yu2024environment,lv2023ris,wang2022ring} & $Q$ & Low & $\boldsymbol{\varphi} \in \{\boldsymbol{\varphi}_1, \boldsymbol{\varphi}_2, \cdots, \boldsymbol{\varphi}_Q\}$ \\
\hline
Proposed & \checkmark & $Q$ & Moderate & $\boldsymbol{\varphi} \in \text{span}\{\boldsymbol{\varphi}_1, \cdots, \boldsymbol{\varphi}_Q\}$ \\
\hline
\end{tabular}}
\begin{tablenotes}
  \footnotesize
  \item CE: Channel estimation \quad PBF: Passive beamforming \quad PC: Predesigned codebook \quad $\text{span}\{\boldsymbol{\varphi}_1, \boldsymbol{\varphi}_2\} = k_1 \boldsymbol{\varphi}_1 + k_2 \boldsymbol{\varphi}_2, k_1, k_2 \in \mathbb{C}$
\end{tablenotes}
\end{table}

However, all the aforementioned codebook schemes focus on the codebook design stage \cite{an2022codebook,an2022Joint,jia2023environment,yu2024environment,lv2023ris,wang2022ring}. For the online codeword configuration stage, they simply select the codeword that yields the optimal performance. As the codebook space is finite, these methods still suffer significant performance gaps compared to the optimal RC configuration. In this letter, we introduce a weighted design operating on codewords of the predesigned codebook for RIS-assisted point-to-point MIMO communications. \emph{In contrast to the direct codeword selection from the predesigned codebook, the proposed scheme designs a set of weights for each codeword in the codebook according to the corresponding output results, yielding a new RIS RC vector.} For illustration, we contrast the proposed scheme to its existing channel estimation, passive beamforming and predesigned codebook counterparts in Table I. The simulation results demonstrate that the proposed scheme has improved performance compared to existing passive beamforming and codebook schemes.

\section{System Model}

We consider an RIS-assisted point-to-point MIMO system in a single cell as shown in Fig. 1, where a base station (BS) with $M_t$ transmit antennas sends $M_s$ data streams to a user with $M_r$ antennas, with $M_s \leq \min\left\{M_t, M_r\right\}$. The RIS consists of $N$ reflecting elements and is equipped with a smart controller capable of adjusting the RCs according to instructions from the BS. The signals from both the cascaded BS-RIS-user link and the direct BS-user link are superimposed at the user. We assume that the frequency-flat baseband equivalent channels spanning from the BS to the RIS, from the RIS to the user, and from the BS to the user are denoted by $\mathbf{H}_t \in \mathbb{C}^{N \times M_t}, \mathbf{H}_r \in \mathbb{C}^{M_r \times N}$ and $\mathbf{H}_d \in \mathbb{C}^{M_r \times M_t}$, respectively.

\begin{figure}[t]
\centering
\includegraphics[width=8.8cm]{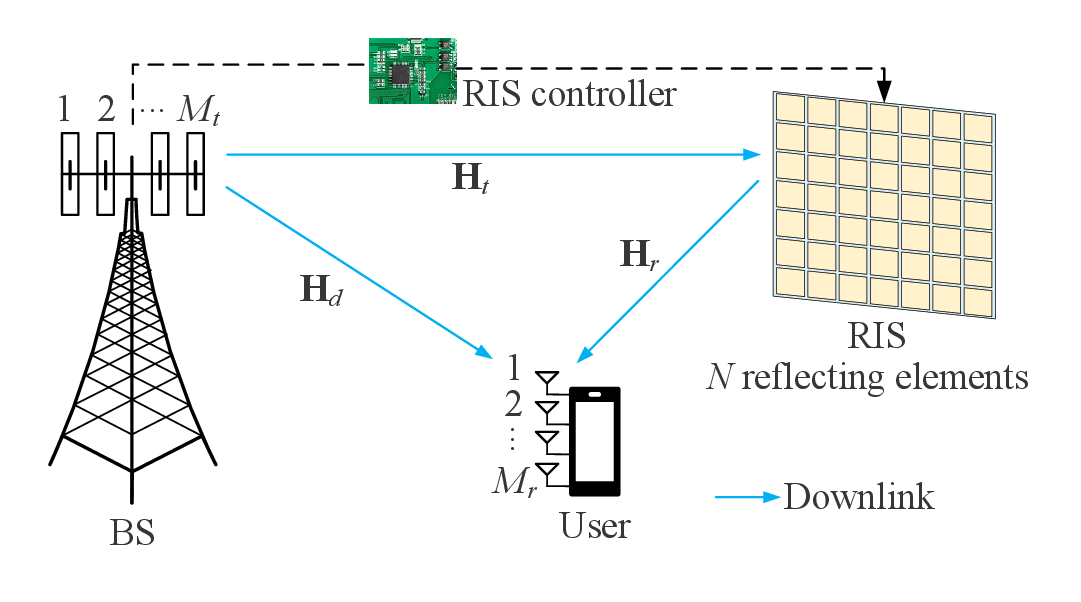}
\caption{An RIS-assisted point-to-point MIMO communication system. }
\label{fig_1}
\end{figure} 

Let $\boldsymbol{\varphi} = [\varphi_1, \varphi_2, \cdots, \varphi_N]^T$ represent the RIS RC vector, where $\varphi_n = e^{j \theta_n}$ denotes the RC of the $n$th RIS element with phase shift $\theta_n$, satisfying $\theta_n \in \left[0, 2\pi \right)$ for $n = 1, 2, \cdots, N$. Thus the composite end-to-end channel $\mathbf{H}_e \in \mathbb{C}^{M_r \times M_t}$ from the BS to the user can be expressed as \begin{align} \mathbf{H}_e = \mathbf{H}_d + \mathbf{H}_r \text{diag}(\boldsymbol{\varphi}) \mathbf{H}_t. \end{align}

During the channel training process in the uplink phase, the user sends the pilot signal $\mathbf{X} = \left[\mathbf{x}_1, \mathbf{x}_2, \cdots, \mathbf{x}_{M_r}\right]^T \in \mathbb{C}^{M_r \times \tau}$ to the BS, where $\mathbf{x}_m^T \in \mathbb{C}^{1 \times \tau}, m = 1, 2, \cdots, M_r$ is the pilot loaded on the $m$th antenna at the user. The pilot matrix satisfies $\lVert \mathbf{X} \rVert_F^2 = \tau p_u$, where $p_u$ is the average pilot power. As we consider a time-division duplexing protocol for both uplink as well as downlink transmissions and assume the channel’s reciprocity, the pilot signal received at the BS is given by \begin{align} \mathbf{Y} = \mathbf{H}_e^H \mathbf{X} + \mathbf{N}_{\text{BS}}, \end{align} where $\mathbf{N}_{\text{BS}} \in \mathbb{C}^{M_t \times \tau}$ denotes the noise matrix at the BS with an average noise power of $\sigma_{\text{BS}}^2$, whose $i$th column vector follows $\mathbf{n}_{\text{BS},i} \sim \mathcal{CN} \left( \mathbf{0}_{M_t}, \sigma_{\text{BS}}^2 \mathbf{I}_{M_t} \right)$, for $i = 1, 2, \cdots, \tau$. We employ mutually orthogonal pilots, and the length of the pilot signal is designed such that $\tau \geq M_r$ \cite{yu2024environment}.

Next, we consider the downlink of data transmission where the BS applies a baseband precoder $\mathbf{W} \in \mathbb{C}^{M_t \times M_s}$ to transmit symbol $\mathbf{s} \in \mathbb{C}^{M_s \times 1}$, with $\mathbb{E}\left\{\mathbf{s} \mathbf{s}^H \right\} = \mathbf{I}_{M_s}$. Furthermore, the precoder satisfies $\lVert \mathbf{W}\rVert_F^2 \leq p_d$ and $p_d$ is the total transmit power at the BS. Thus the received signal at the user is obtained as \begin{align} \mathbf{r} = \mathbf{H}_e \mathbf{W} \mathbf{s} + \mathbf{n}_{\text{UE}}, \end{align} where $\mathbf{n}_{\text{UE}} \in \mathbb{C}^{M_r \times 1}$ is the noise at the user with an average noise power of $\sigma_{\text{UE}}^2$, satisfying $\mathbf{n}_{\text{UE}} \sim \mathcal{CN} \left(\mathbf{0}_{M_r}, \sigma_{\text{UE}}^2\mathbf{I}_{M_r} \right)$.

Meanwhile, we adopt the Rician channel in this paper. Specifically, the RIS-user channel can be expressed as 
\begin{align} \label{eq:5} \mathbf{H}_{r} = \sqrt{\beta_r} \left( \sqrt{\frac{F_r}{F_r + 1}}\mathbf{H}_{r}^{\text{LoS}} + \sqrt{\frac{1}{F_r + 1}}\mathbf{H}_{r}^{\text{NLoS}} \right), \end{align} 
where $\beta_r$ and $F_r$ are the path loss and the Rician factor of RIS-user channel, respectively; $\mathbf{H}_{r}^{\text{LoS}} \in \mathbb{C}^{M_r \times N}$ and $\mathbf{H}_{r}^{\text{NLoS}} \in \mathbb{C}^{M_r \times N}$ represent the line-of-sight (LoS) and the non-line-of-sight (NLoS) components of the RIS-user channel, respectively. The element on the $m_r$th row and the $n$th column of the NLoS matrix is modeled by Rayleigh fading, which follows ${\mathbf{H}_{r}^{\text{NLoS}}}_{m_r, n} \sim \mathcal{CN}\left(0, 1\right)$. Similarly, the BS-user channel and BS-RIS channel can be modeled by using (\ref{eq:5}).

Moreover, we consider a uniform linear array (ULA) at the BS, a ULA at the user, and a uniform planar array (UPA) at the RIS. Let $\mathbf{a}_{\text{BS}}\left(\delta\right) \in \mathbb{C}^{M_t \times 1}$, $\mathbf{a}_{\text{UE}}\left(\delta\right) \in \mathbb{C}^{M_r \times 1}$ and $\mathbf{a}_{\text{R}}\left(\zeta, \gamma\right) \in \mathbb{C}^{N \times 1}$ denote the steering vector of the BS, the user and the RIS, respectively. Specifically, the $m_t$th entry of $\mathbf{a}_{\text{BS}}$ is denoted as $e^{j \frac{2 \pi}{\lambda} (m_t-1) d_{\text{BS}} \sin(\delta) }, m_t = 1, 2, \cdots, M_t$, where $d_{\text{BS}}$ denotes the element spacing of the BS, $\lambda$ denotes the signal wavelength, and $\delta \in \left[-\pi/2, \pi/2\right)$ denotes the angle of departure (AoD) or the angle of arrival (AoA). Similarly, the $m_r$th entry of $\mathbf{a}_{\text{UE}}$ is denoted as $e^{j \frac{2 \pi}{\lambda} (m_r-1) d_{\text{UE}} \sin(\delta) }, m_r = 1, 2, \cdots, M_r$, where $d_{\text{UE}}$ denotes the element spacing of the user. The $n$th entry of $\mathbf{a}_{\text{R}}$ is denoted as $e^{j 2 \pi d_{\text{R}} \sin(\gamma) \left[ \lfloor \frac{n-1}{N_x} \rfloor \sin(\zeta) + ((n-1)-\lfloor \frac{n-1}{N_x} \rfloor N_x) \cos(\zeta) \right]/\lambda }, n = 1, 2, \cdots, N$, where $d_{\text{R}}$ denotes the element spacing of the RIS. $N_x$ is the number of elements deployed at each row of the RIS. $\zeta \in \left[0, \pi\right)$ and $\gamma \in \left[-\pi/2, \pi/2\right)$ denote the azimuth and elevation AoA/AoD, respectively. Thus, the LoS component of the $\mathbf{H}_t$, $\mathbf{H}_r$ and $\mathbf{H}_d$ are given by $\mathbf{a}_{\text{R}}\left(\zeta_t^{\text{AoA}}, \gamma_t^{\text{AoA}}\right) \mathbf{a}_{\text{BS}}\left(\delta_t^{\text{AoD}}\right)^H$, $\mathbf{a}_{\text{UE}}\left(\delta_r^{\text{AoA}}\right) \mathbf{a}_{\text{R}}\left(\zeta_r^{\text{AoD}}, \gamma_r^{\text{AoD}}\right)^H$ and $\mathbf{a}_{\text{UE}}\left(\delta_d^{\text{AoA}}\right) \mathbf{a}_{\text{BS}}\left(\delta_d^{\text{AoD}}\right)^H$, respectively, where $\delta_t^{\text{AoD}}$, $\zeta_t^{\text{AoA}}$ and $\gamma_t^{\text{AoA}}$ represent the AoD, the azimuth and elevation AoA from the BS to the RIS, respectively; $\delta_r^{\text{AoA}}$, $\zeta_r^{\text{AoD}}$ and $\gamma_r^{\text{AoD}}$ represent the AoA, the azimuth and elevation AoD from the RIS to the user, respectively; $\delta_d^{\text{AoA}}$ and $\delta_d^{\text{AoD}}$ represent the AoA and the AoD from the BS to the user, respectively.

In the next section, unlike traditional passive beamforming and codebook schemes, the proposed scheme maximizes the channel capacity of the point-to-point MIMO systems by leveraging online codebook weighting design.

\section{The Proposed Codebook Weighting Scheme}
\begin{figure}[t]
\centering
\includegraphics[width=8.8cm]{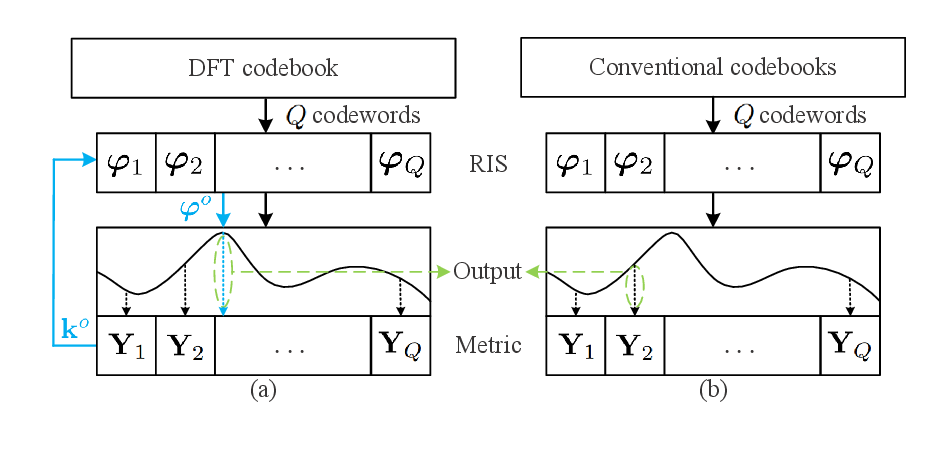}
\caption{Comparison between (a) The proposed DFT codebook weighting scheme and (b) The conventional codebook scheme.}
\label{fig_framework}
\end{figure}

In this section, we propose a codebook weighting scheme, and compare it with conventional codebook schemes in Fig. \ref{fig_framework}. Specifically, during the online RIS RC configuration stage, conventional codebooks choose the codeword resulting in optimal performance from the predesigned codewords, i.e. $\boldsymbol{\varphi}_{\hat{q}}$, while the proposed scheme leverages the observations of these codewords through the uplink phase to design a new RIS RC vector $\boldsymbol{\varphi}^o$, which achieves better performance.

Denote the $(N+1)$-dimensional DFT matrix as $\mathbf{A}$, i.e., $\mathbf{A}_{m,n} = e^{-j \frac{2\pi m n}{N+1}}, m = 0, 1, \cdots, N, n = 0, 1, \cdots, N$. The $q$th codeword in the DFT codebook is composed of the elements from the $2$nd to the $(N + 1)$th entries in the $q$th column of $\mathbf{A}$, with $q = 1, 2, \cdots, N+1$. During the $q$th training block, we configure the RIS using the $q$th codeword. Thus the signal received at the BS in the uplink phase is given by \begin{align} \mathbf{Y}_q = \left( \mathbf{H}_d^H + \mathbf{H}_t^H \text{diag}(\boldsymbol{\varphi}_q^H) \mathbf{H}_r^H\right) \mathbf{X} + \mathbf{N}_{\text{BS}, q}, \end{align} where $\mathbf{N}_{\text{BS}, q}$ is the noise at the BS when configuring the $q$th codeword. Let $\mathbf{H}_{c,n}^H = \mathbf{h}_{t,n} \mathbf{h}_{r,n}^H \in \mathbb{C}^{M_t \times M_r}, n = 1, 2, \cdots, N$ represent the cascaded channel via the $n$th RIS element, where $\mathbf{h}_{t,n}$ and $\mathbf{h}_{r,n}^H$ denote the $n$th column of $\mathbf{H}_t^H$ and the $n$th row of $\mathbf{H}_r^H$, respectively. After collecting signals across $Q$ training blocks, we can obtain that 
\begin{align}
\Breve{\mathbf{Y}} = \left( \mathbf{A}_Q^H \otimes \mathbf{I}_{M_t} \right) \mathbf{H} \mathbf{X}
+ \Breve{\mathbf{N}},
\end{align} where $\Breve{\mathbf{Y}} = \left[\mathbf{Y}_{1}^H, \mathbf{Y}_{2}^H, \cdots, \mathbf{Y}_{Q}^H \right]^H \in \mathbb{C}^{QM_t \times \tau}$ are the received signals, $\Breve{\mathbf{N}} = \left[\mathbf{N}_{\text{BS},1}^H, \mathbf{N}_{\text{BS},2}^H, \cdots, \mathbf{N}_{\text{BS},Q}^H \right]^H \in \mathbb{C}^{QM_t \times \tau}$, $\mathbf{H} = [\mathbf{H}_{d}, \mathbf{H}_{c,1}, \cdots, \mathbf{H}_{c,N}]^H \in \mathbb{C}^{(N+1)M_t \times M_r}$, and $\mathbf{A}_Q \in \mathbb{C}^{(N+1) \times Q}$ collects the $Q$ columns of the DFT matrix $\mathbf{A}$.

Next, we consider two scenarios with different training overhead:
\paragraph{$Q < N+1$} By adopting the minimum norm solution, the estimate of $\mathbf{H}$ can be derived as
\begin{align} \label{Q_1}
\hat{\mathbf{H}} = \left( \mathbf{A}_Q \left(\mathbf{A}_Q^H \mathbf{A}_Q\right)^{-1} \otimes \mathbf{I}_{M_t} \right) \Breve{\mathbf{Y}} \mathbf{X}^H \left(\mathbf{X} \mathbf{X}^H\right)^{-1},
\end{align} where $\hat{\mathbf{H}} = [\hat{\mathbf{H}}_{d}, \hat{\mathbf{H}}_{c,1}, \cdots, \hat{\mathbf{H}}_{c,N}]^H$ represents the estimate of the direct and $N$ cascaded channels.

\paragraph{$Q = N+1$} (\ref{Q_1}) can be simplified to 
\begin{align} \label{Q_2}
\hat{\mathbf{H}} = \frac{1}{N+1} \left(\mathbf{A}^H \otimes \mathbf{I}_{M_t} \right) \Breve{\mathbf{Y}} \mathbf{X}^H \left(\mathbf{X} \mathbf{X}^H\right)^{-1}.
\end{align}

Then, in order to obtain the weight vector and the transmit precoding matrix, we formulate a MIMO channel capacity maximization problem, yielding \begin{align} \label{P_1} \max_{\mathbf{k}, \mathbf{W}} C = & \text{log}_2 \left| \mathbf{I}_{M_r} + \frac{1}{\sigma_{\text{UE}}^2} \mathbf{H}^H \left(\mathbf{A}_Q \mathbf{k} \otimes \mathbf{W} \right) \left(\mathbf{A}_Q \mathbf{k} \otimes \mathbf{W} \right)^H \mathbf{H} \right| \notag \\ \mathrm{s.t.} & \left|\mathbf{a}_{Q,n}^H \mathbf{k} \right| = 1, n = 1, 2, \cdots, N+1, \notag \\ & \lVert \mathbf{W} \rVert_F^2 \leq p_d, \end{align} where $\mathbf{k} = [k_1, k_2, \cdots, k_{Q}]^T$ and $k_q$ is the $q$th weight coefficient corresponding to the $q$th codeword in the DFT codebook. $\mathbf{a}_{Q,n}^H \in \mathbb{C}^{1 \times Q}$ denotes the $n$th row of $\mathbf{A}_Q$.

Considering the non-convex constraint of the RIS RCs, it is hard to obtain the optimal weight vector to the problem (\ref{P_1}). To solve this problem, we first initialize the weight vector $\mathbf{k} = \left(\mathbf{A}_Q^H \mathbf{A}_Q\right)^{-1} \mathbf{A}_Q^H \boldsymbol{\varphi}_m$, where $\boldsymbol{\varphi}_m$ is the codeword yielding the maximum channel capacity during the observations of $Q$ training blocks. Next, we iteratively optimize the weight coefficient vector and obtain the transmit precoding matrix \cite{WCL_2024_Lin_Stacked, WCL_2025_Huang_Stacked}.

\subsubsection{Weight Design}
To find a near-optimal solution, we perform the singular value decomposition (SVD) of $\mathbf{H}^H = \mathbf{U} \mathbf{\Lambda} \mathbf{V}^H$, where $\mathbf{V} \in \mathbb{C}^{(N+1) M_t \times \text{rank}\left(\mathbf{H}^H \right)}$. Then we construct $\mathbf{P}$ by collecting the first $M_s$ columns of $\mathbf{V}$, yielding \begin{align} \mathbf{P} = \sqrt{\frac{N+1}{M_S}} \mathbf{V}_{:,1:M_s} = \left[\mathbf{P}_1^T, \mathbf{P}_2^T, \cdots, \mathbf{P}_{N+1}^T \right]^T.\end{align} 

According to \cite{zhou2020joint}, the problem in (\ref{P_1}) can be simplified by maximizing the lower bound of the MIMO channel capacity instead. And its optimality can be readily proven by using \emph{Lemma 2} and \emph{Proposition 6} in \cite{zhou2020joint}. As a result, we have \begin{align} \label{P_k} \max_{\mathbf{k}} & \, \mathbf{k}^H \mathbf{A}_Q^H \mathbf{B} \mathbf{A}_Q \mathbf{k} \notag \\ \mathrm{s.t.} & \left|\mathbf{a}_{Q,n}^H \mathbf{k} \right| = 1, n = 1, 2, \cdots, N+1, \end{align} where $\mathbf{B} \in \mathbb{C}^{(N+1) \times (N+1)}$ is a Hermitian matrix whose elements satisfy $\mathbf{B}_{i,j} = \text{trace}\left(\mathbf{P}_j^H \mathbf{P}_i \right)$.

Notice that the Karush-Kuhn-Tucker (KKT) condition of (\ref{P_k}) can be derived following the monotone convergence theorem, which can be expressed as \begin{align} \label{KKT}
    \mathbf{B} \mathbf{A}_Q \mathbf{k} - \sum_{n = 1}^{N+1} \upsilon_n \mathbf{e}_n \mathbf{a}_{Q,n}^H \mathbf{k} = \mathbf{0},
\end{align} where $\upsilon_n$ and $\mathbf{e}_n$ are the $n$th Lagrange multiplier and standard unit vector, respectively.

Next, we adopt an iterative method to optimize the weight vector. Specifically, in order to satisfy the KKT condition, given the weight vector $\mathbf{k}^r$ of the $r$th iteration, the weight vector and Lagrange multiplier vector $\boldsymbol{\upsilon}^{r+1} = \left[\upsilon_1^{r+1}, \upsilon_2^{r+1}, \cdots, \upsilon_{N+1}^{r+1} \right]^T$ of the $(r+1)$th iteration can be obtained by
\begin{align} \label{k,lambda} \mathbf{A}_Q \mathbf{k}^{r+1} &= e^{j \angle(\mathbf{B} \mathbf{A}_Q \mathbf{k}^r)}, \notag \\ \boldsymbol{\upsilon}^{r+1} &= \left| \mathbf{B} \mathbf{A}_Q \mathbf{k}^r \right|. \end{align}
We iteratively optimize the weight coefficient vector until the norm of $\mathbf{k}$ tends to converge or the number of iterations reaches a preset maximum value $r_{\text{max}}$.

\subsubsection{Transmit Precoding Design}
Given the optimized weight coefficient vector $\mathbf{k}$, the optimization problem in (\ref{P_1}) can be reformulated as
\begin{align} \label{eq:W}
    \max_{\mathbf{W}} & \, \text{log}_2 \left| \mathbf{I}_{M_r} + \frac{1}{\sigma_{\text{UE}}^2} \mathbf{H}_e \mathbf{W} \mathbf{W}^H \mathbf{H}_e^H \right| \notag \\ \mathrm{s.t.} & \lVert \mathbf{W} \rVert_F^2 \leq p_d,
\end{align}
where $\mathbf{H}_e = \mathbf{H}^H \left(\mathbf{A}_Q \mathbf{k} \otimes \mathbf{I}_{M_t} \right)$ denotes the downlink composite channel. Note that the problem formulated in (\ref{eq:W}) turns out to be a convex optimization problem that can be solved through SVD transmission \cite{arXiv_2024_Jiancheng_Emerging, el2014spatially}. 

Specifically, denote $\mathbf{H}_e = \Tilde{\mathbf{U}} \Tilde{\mathbf{\Lambda}} \Tilde{\mathbf{V}}^H$ as the truncated SVD of $\mathbf{H}_e$, where $\Tilde{\mathbf{V}} \in \mathbb{C}^{M_t \times M_s}$. The optimal $\mathbf{W}$ is given by
\begin{align} \label{W} \mathbf{W} = \Tilde{\mathbf{V}} \text{diag}\left(p_1, p_2, \cdots, p_{M_s} \right)^{\frac{1}{2}}, \end{align} where $p_i$ represents the power allocated to the $i$th data stream determined by using classic water-filling power allocation algorithm. Specifically, the transmit power allocated to the $i$th data stream is given by $p_i = \max \left(1/\eta - \sigma_{\text{UE}}^2/\Tilde{\mathbf{\Lambda}}_{i, i}^2, 0 \right),$ where $\eta$ is a threshold that satisfies $\sum_{i=1}^{M_s} p_i = p_d$.

After performing the proposed algorithm in (\ref{k,lambda}) and (\ref{W}), we can obtain the optimal weight coefficient vector $\mathbf{k}^o = \left[k_1^o, k_2^o, \cdots, k_{Q}^o \right]^T$ and its corresponding transmit precoding matrix $\mathbf{W}^o$. Then, during the downlink data transmission phase, we adopt the optimal weight coefficient and thus the optimal RIS RC vector can be calculated by $\boldsymbol{\varphi}^o = \sum_{q=1}^{Q} k_q^o \boldsymbol{\varphi}_q.$

Note that the total computational complexity of the proposed codebook weighting design is predominantly given by $\mathcal{O}\left(N^2 M_t M_s^2 + Q^2 r_{\text{max}}\right)$, which includes computing the matrix $\mathbf{B}$ and solving the KKT condition iteratively. Compared to existing codebook schemes, the proposed scheme eliminates the non-negligible performance loss caused by finite predesigned codebook space at the cost of moderate complexity.

\section{Codeword Configuration Order}
In this section, we analyze the effect of the codeword configuration order on the performance of the proposed scheme.

Firstly, we apply the sequential codeword configuration order which is applicable under all channel conditions without any prior. However, when the training overhead is small, the channel information obtained from codeword configurations is limited, potentially leading to significant errors in the design of codeword weights. To reduce this adverse effect, we utilize the channel statistics to optimize the codeword configuration order for further enhancing system performance. Specifically, these two methods are summarized below:
\begin{itemize}
\item[1)] In the absence of prior information about the channels, we sequentially select codewords. Specifically, based on the training overhead $Q$, we choose the first $Q$ codewords from the DFT codebook to configure RIS RC in turn.
\item[2)] With the knowledge of the LoS components of the channels based on statistical CSI, we employ an environment-aware codeword configuration order. Specifically, the codeword that aligns the LoS components to the maximum extent is prioritized to configure the RIS RC, since it is statistically more likely to lead to good performance.
\end{itemize}

To further explain the proposed environment-aware codeword configuration method, we represent the LoS components of the RIS-user, BS-RIS and BS-user channels as $\mathbf{H}_r^{\text{LoS}}$, $\mathbf{H}_t^{\text{LoS}}$ and $\mathbf{H}_d^{\text{LoS}}$, respectively. The configuration order of codewords is determined according to their alignment degree to the LoS component. For simplicity, we choose an arbitrary antenna pair at the transmitter and receiver, indexed by $m_t$ and $m_r$, respectively \cite{xu2023antenna}. As a result, the codeword configuration order is determined by evaluating \begin{align} \label{q_order} L_q = \left|\mathbf{H}_{d,m_r,m_t}^{\text{LoS}} + \mathbf{H}_{r,m_r}^{\text{LoS}} \text{diag}\left(\boldsymbol{\varphi}_q \right) \mathbf{H}_{t,m_t}^{\text{LoS}} \right|^2. \end{align} 

We first calculate the metrics $L_q, q = 1, 2, \cdots, N+1$ based on (\ref{q_order}) for all codewords in the codebook, then $Q$ codewords resulting in the highest channel gain are selected for RIS RC configuration based on the determined training overhead $Q$.

\section{Simulation Results}
\begin{figure}[!t]    
\centering            
\subfloat[\scriptsize Perfect CSI]  
{
\label{fig_Q:subfig1}\includegraphics[width=0.45\textwidth]{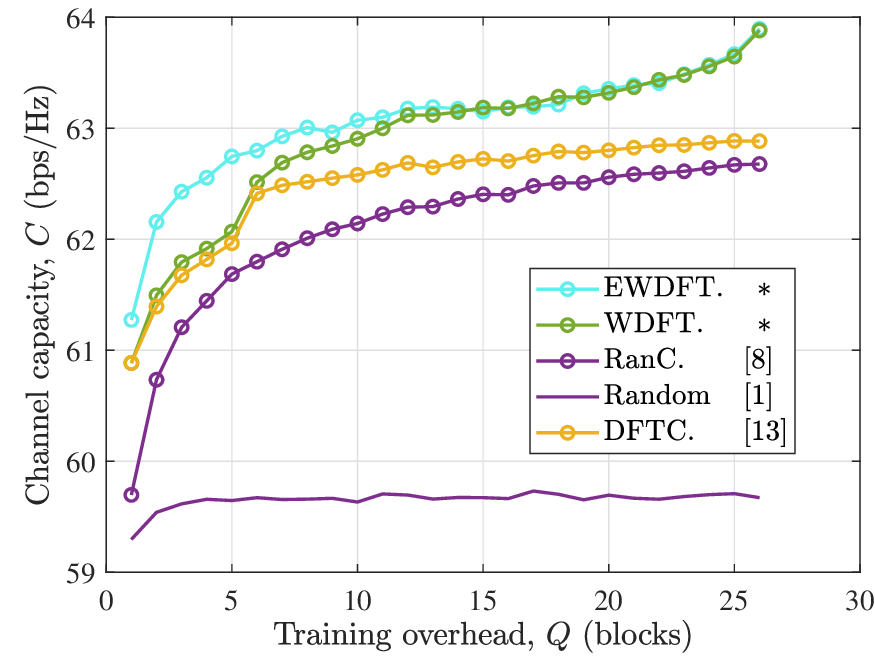}
}\\
\subfloat[\scriptsize Imperfect CSI]
{
\label{fig_Q:subfig2}\includegraphics[width=0.45\textwidth]{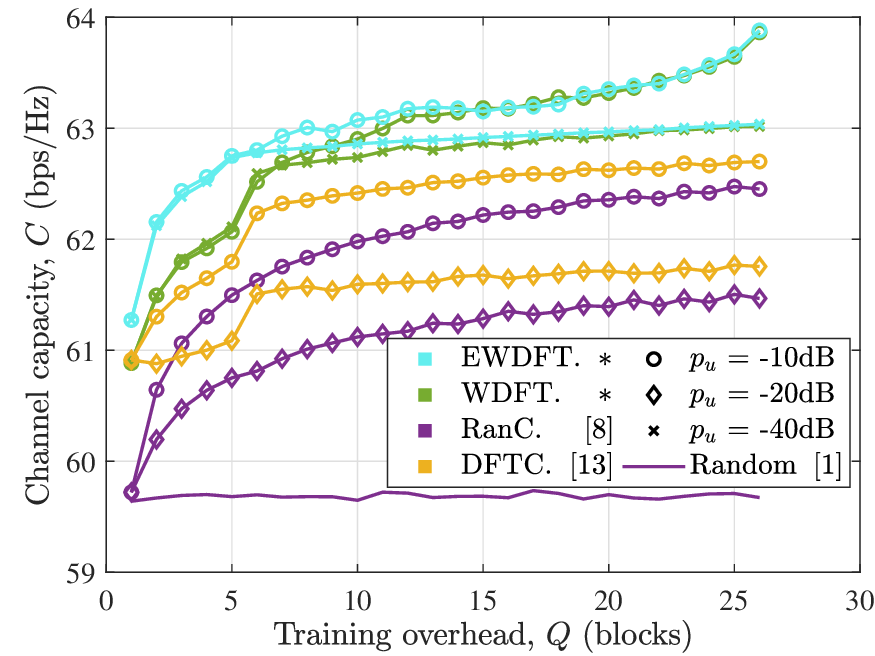}
}
\caption{Channel capacity $C$ versus training overhead $Q$.}   
\label{fig_Q}       
\end{figure}

In this section, we provide simulation results to validate the performance of the proposed scheme. All results are obtained by averaging 1,000 independent experiments. We consider a 3D Cartesian coordinate system, where the antenna array at the BS and the user are modeled by a ULA with antenna spacing of $d_{\text{BS}} = d_{\text{UE}} = \lambda/2$. The number of the BS antennas and the user antennas are set to $M_t = 4$ and $M_r = 4$, respectively. The RIS is modeled by a UPA with $5 \times 5 = 25$ elements. The element spacing is set to $d_{\text{R}} = \lambda/4$ \cite{an2021low}. The locations of the reference antenna/element at the BS, the RIS, and the user are set as (0, 0, $h_{\text{BS}}$), (0, $d_{\text{BR}}$, $h_{\text{R}}$) and ($d_{\text{U}}$, $d_{\text{BR}}$, 0), respectively. Specifically, the height of the BS, the height of the RIS, the distance from the user to $y\text{-}z$ plane and the distance between BS and RIS are $h_{\text{BS}} = \SI{5}{m}$, $h_{\text{R}} = \SI{5}{m}$, $d_{\text{U}} = \SI{3}{m}$ and $d_{\text{BR}} = \SI{100}{m}$, respectively. The Rician factors of $\mathbf{H}_t$, $\mathbf{H}_r$ and $\mathbf{H}_d$ are $F_t = \SI{6}{dB}$, $F_r = \SI{4}{dB}$ and $F_d = \SI{3}{dB}$, respectively \cite{yu2024environment}. The path loss of each channel is modeled as $\beta = C_0(d/d_0)^{-\alpha}$, where $d$ is the distance of corresponding link, $C_0 = \SI{-20}{dB}$ denotes the path loss of the reference distance $d_0 = \SI{1}{m}$ and $\alpha$ denotes the path loss factor \cite{an2021low}. The path loss factors of $\mathbf{H}_t$, $\mathbf{H}_r$ and $\mathbf{H}_d$ are $\alpha_t = 2.4$, $\alpha_r = 2.5$ and $\alpha_d = 3.5$, respectively. Moreover, the transmit power at the BS is set to $p_d = \SI{30}{dBm}$, the average noise power at the BS and the user are $\sigma_{\text{BS}}^2 = \SI{-120}{dBm}$ and $\sigma_{\text{UE}}^2 = \SI{-110}{dBm}$, respectively. The average power of the pilot signals is shown in Fig. \ref{fig_Q}(b). In this context, we consider the proposed DFT codebook weighting (WDFT.) scheme and four benchmark schemes, including the random phase shift (Random) scheme \cite{an2021low}, random codebook (RanC.) scheme \cite{an2022codebook}, DFT codebook (DFTC.) scheme \cite{wang2022ring} and passive beamforming optimization scheme (CE \& PBF.) with grouping strategy \cite{kundu2022optimal}. Besides, the proposed scheme with the environment-aware codeword configuration order is marked as EWDFT. 

Firstly, Fig. \ref{fig_Q} evaluates the MIMO channel capacity versus the training overhead. To evaluate the performance upper bound of the proposed scheme, we assume that each training block in Fig. \ref{fig_Q}(a) is devoid of noise. It can be seen that as the training overhead $Q$ increases, the proposed scheme gets performance improvement in terms of the channel capacity and always outperforms the DFT codebook scheme as it utilizes the optimal codeword obtained in the DFT codebook to initialize the weight vector. As the training overhead increases, the system performance of the random codebook and DFT codebook schemes improves but experiences diminishing returns. This is because they only select the codeword that performs best from the predesigned codebook. By contrast, the proposed weighted scheme can progressively traverse the whole solution space according to multiple orthogonal codewords. Furthermore, considering the two codeword configuration orders proposed in Section IV, it can be observed from Fig. \ref{fig_Q}(a) that the environment-aware codeword configuration order can significantly enhance system performance for a relatively low training overhead, owing to its utilization of the statistical CSI to prioritize codewords that align with the LoS component. As the training overhead increases, the advantages of codeword selection gradually diminish.

\begin{figure}[!t]    
\centering            
\subfloat[\scriptsize]  
{
\label{fig_pd_N:subfig1}\includegraphics[width=0.45\textwidth]{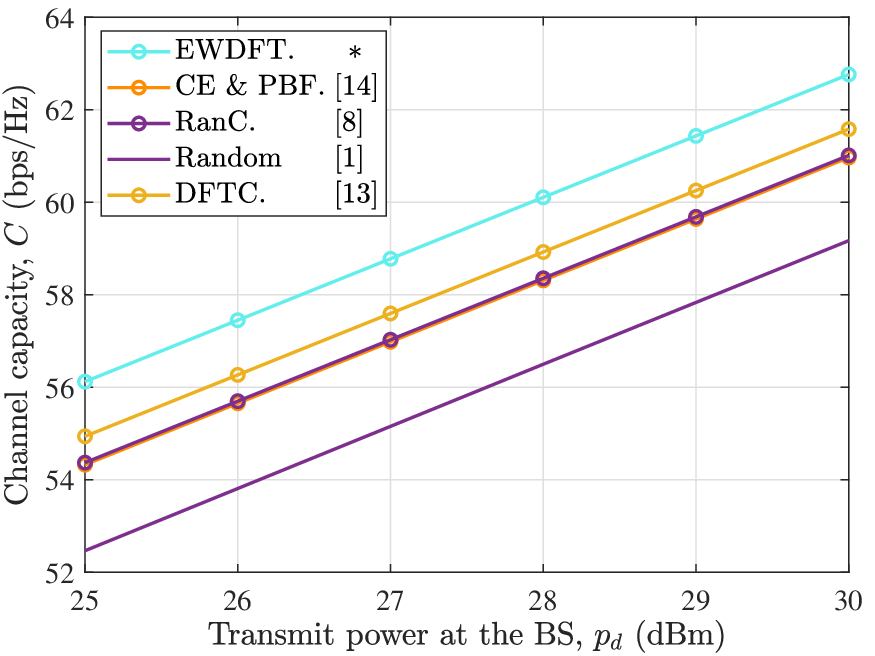}
}\\
\subfloat[\scriptsize]
{
\label{fig_pd_N:subfig2}\includegraphics[width=0.45\textwidth]{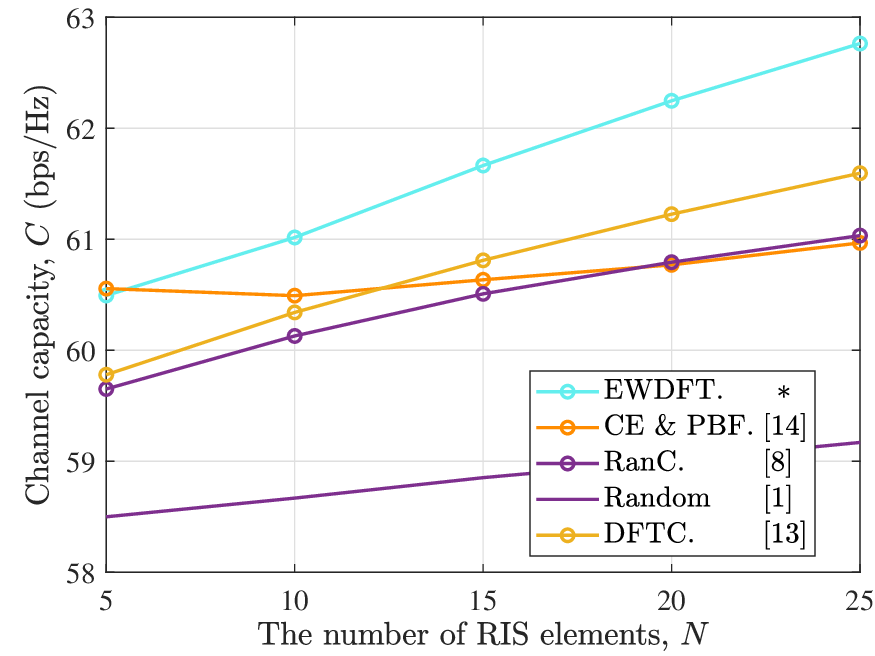}
}
\caption{(a) Channel capacity $C$ versus total transmit power $p_d$. (b) Channel capacity $C$ versus the number of RIS elements $N$. (Both with training overhead $Q = 6$.)}  
\label{fig_pd_N}  
\end{figure}

Besides, Fig. \ref{fig_Q}(b) evaluates the effect of the channel estimation errors on the proposed codebook weighting scheme. We observe that as the pilot power reduces, the proposed scheme exhibits a smaller performance degradation compared to the random and DFT codebook schemes, indicating its stronger robustness against noise. Note that for a larger training overhead, the impact of channel estimation errors becomes more significant. As a result, the performance loss of the proposed scheme is nearly negligible at low training overhead, but becomes more pronounced as the training overhead increases.

Then, Fig. \ref{fig_pd_N}(a) examines the channel capacity versus the total transmit power, where the training overhead is set to $Q = 6$. We can observe a logarithmic increase relationship between the channel capacity and the transmit power at the BS. Again, the proposed scheme outperforms the existing codebook and passive beamforming schemes under all setups considered. Next, Fig. \ref{fig_pd_N}(b) evaluates the channel capacity versus the number of RIS elements $N$, where we set the transmit power $p_d = \SI{30}{dBm}$. It can be seen that with the increase of the RIS elements $N$, the channel capacity of the proposed scheme increases, which is due to the fact that a larger number of reflecting elements could reap more radiating energy by coherently superimposing all cascaded links. Although the passive beamforming scheme can achieve the optimal channel capacity for a small number of RIS elements (e.g., $N = 5$), training overhead becomes insufficient as $N$ increases, resulting in a restricted performance improvement with respect to large scale $N$. Furthermore, we observe that the performance gain of the proposed scheme over other codebook schemes becomes more significant for a large-size RIS. This is attributed to the superiority of the RIS RC solution obtained through the optimization of the weight coefficients and transmit precoding matrix.

\section{Conclusion}
In this paper, we proposed a DFT codebook weighting design for point-to-point MIMO systems. Specifically, the proposed scheme configured the RIS RC based on codewords from the DFT codebook and collected the received signals at the BS for all training blocks. Then, we utilized these observations to design the weight coefficient vector and the corresponding transmit precoding matrix. Finally, the optimal RIS RC vector was obtained by calculating the weighted sum of the orthogonal codewords. Furthermore, we considered the impact of different codeword configuration orders on the performance and evaluated the channel capacity of the DFT codebook weighting scheme through simulation results.

\bibliographystyle{IEEEtran}
\bibliography{IEEEabrv,Ref}

\end{document}